\documentclass{article}

\usepackage{microtype}
\usepackage{graphicx}
\usepackage{subfigure}
\usepackage{booktabs} 
\usepackage{hyperref}
\usepackage{amsmath, eqparbox}
\usepackage{amssymb}
\usepackage{apacite}

\usepackage[accepted]{icml2021}

\icmltitlerunning{Practical considerations on using private sampling for synthetic data}
\usepackage{amsthm}
\usepackage{optidef}
\newtheorem{theorem}{Theorem}[section]

\theoremstyle{definition}
\newtheorem{definition}{Definition}[section]
\theoremstyle{remark}

\usepackage{mathtools, stmaryrd}
\usepackage{xparse} \DeclarePairedDelimiterX{\Iintv}[1]{\llbracket}{\rrbracket}{\iintvargs{#1}}
\NewDocumentCommand{\iintvargs}{>{\SplitArgument{1}{,}}m}
{\iintvargsaux#1} %
\NewDocumentCommand{\iintvargsaux}{mm} {#1\mkern1.5mu,\mkern1.5mu#2}
\DeclareMathOperator*{\argmin}{arg\,min}
\setcitestyle{square}

\begin{document}

\twocolumn[
\icmltitle{Practical considerations on using private sampling for synthetic data}



\icmlsetsymbol{equal}{*}

\begin{icmlauthorlist}
\icmlauthor{Clément Pierquin}{craft}
\icmlauthor{Bastien Zimmermann}{craft}
\icmlauthor{Matthieu Boussard}{craft}
\end{icmlauthorlist}

\icmlaffiliation{craft}{Craft.AI, Paris, FRA}

\icmlcorrespondingauthor{Clément Pierquin}{Clement.Pierquin@craft.ai}
\icmlcorrespondingauthor{Bastien Zimmermann}{Bastien.Zimmermannn@craft.ai}
\icmlcorrespondingauthor{Matthieu Boussard}{Matthieu.Boussard@craft.ai}

\icmlkeywords{Machine Learning, ICML}

\vskip 0.3in
]



\printAffiliationsAndNotice{}

\begin{abstract}
Artificial intelligence and data access are already mainstream. One of the main challenges when designing an artificial intelligence or disclosing content from a database is preserving the privacy of individuals who participate in the process. Differential privacy for synthetic data generation has received much attention due to the ability of preserving privacy while freely using the synthetic data. Private sampling is the first noise-free method to construct differentially private synthetic data with rigorous bounds for privacy and accuracy. However, this synthetic data generation method comes with constraints which seem unrealistic and not applicable for real-world datasets. In this paper, we provide an implementation of the private sampling algorithm and discuss the realism of its constraints in practical cases.

\end{abstract}

\section{Introduction}
Recently, there have been more and more regulations in Europe about privacy \cite{dataAct}. In the context of IoT for example, databases must be protected to ensure privacy protection for the users. Several ideas have already been explored in order to privatize data, such as de-identification or k-anonymization \cite{KANO}.
Today, differential privacy \cite{DP} (DP) is the gold standard for privacy protection of randomized algorithms. Its main objective is to prevent an adversary from getting access to a particular element in a dataset while releasing useful statistical information about the database. Standards DP algorithms only work for specified queries performed on a dataset and therefore lack of flexibility.

Synthetic data, already used for data augmentation, can be used for data protection. It is more flexible than standard DP algorithms since once synthetic data is generated, it ideally can be used or disclosed without privacy concerns. In this case, synthetic data generation requires privacy-preserving mechanisms to get rigorous privacy guarantees. Let us take the example of generative adversarial networks (GANs). GANs are not reversible so it is not possible to make a deterministic function go from the generated samples to the real samples. In fact, the generated samples are created using an implicit distribution of the real data. However, one cannot rely on the fact that GANs are not reversible to simply naively use them and hope that the system is privacy-preserving \cite{GANINF}.

The most simple way to generate differentially private synthetic data is by adding Laplacian noise on each point in the dataset. However, there is a great loss in utility during this process and the synthetic data cannot be used, even for simple queries. More generally, when privately training a deep learning model, there is a tradeoff between privacy and utility \cite{NF}. This is why machine learning algorithms attain lower performances with DP synthetic data as a training set than with the original data. While the most used methods to generate DP synthetic data are deep neural network models based, some partly statistical or Bayesian methods can be used in particular cases and outperform usual models. For example, the 2018 NIST differential privacy synthetic data competition has been won by a graphical model \cite{NIST}.

Private sampling \cite{PS}, the method studied in this work, is a method for synthetic data generation which, under certain assumptions, proposes the first noise-free method to construct differentially private synthetic data. In the light of the outstanding offered guarantees, it is interesting to study this method thoroughly. The low-dimensional marginals of the constructed density exactly match the empirical density ones, and privacy is controlled with rigorous bounds. However, these bounds seem too restrictive for the algorithm to be used in practice. In this paper, our main contribution is a discussion about the realism of the constraints which must be respected in order to get differentially private data. We also provide the code\footnote{https://github.com/craft-ai/private-sampling} and the linear and quadratic programs (QP) needed to compute efficiently the density of the private sampling algorithm. 

\section{Related work}

In this section, we present the mathematical framework behind private sampling introduced in \cite{PS}. First, we recall the definition of differential privacy.
\begin{definition}[Differential Privacy]
A randomized algorithm $\mathcal{M} \colon S^N \mapsto \mathcal{R}$ satisfies $\varepsilon$-differential privacy if for any two adjacent datasets $X_1, X_2 \in S^N$ differing by one element, and any output subset $\mathcal{O} \subset \mathcal{R}$ it holds :\[P \left[\mathcal{M}(X_1)\in \mathcal{O}\right] \leq e^\varepsilon P \left[\mathcal{M}(X_2)\in \mathcal{O} \right]\]
\end{definition}
Let $n, p, k, m \in \mathbb{N}^*$.
Consider  $X = (x_1, . . . , x_n)$ a sample from the Boolean cube $\{-1, 1\}^p$. We want to generate $Y = (y_1, . . . , y_k) \subset \{-1, 1\}^p$ from a density $h$ such that the randomized algorithm taking $X$ and other parameters as input and returning $Y$ is differentially private. Moreover, in order to address the privacy-utility issue, 
we want $h$ to have the same low-dimensional marginals as $f_n$, the empirical distribution of $X$. The core ingredient for marginal matching is Fourier analysis on the Boolean cube. 
The Walsh functions $w_J$ on subsets $J \subset[p]$ are defined by : 
$w_J \colon \{-1,1\}^p \to \mathbb{R},$\\
  $\forall x\in \{-1,1\}^p, w_J(x) = \prod_{j\in J}x(j)$ with $w_\emptyset = 1$.\\
We introduce the canonical inner product of $L^2(\{-1,1\}^p)$ : $\langle f, g\rangle = \sum_{x \in \{-1,1\}^p}f(x)g(x)$.\\Let $f \colon \{-1,1\}^p \to \mathbb{R}$. Then, we have the following results:
\begin{itemize}
\item \eqmakebox{$(w_J)_{J \subset[p]}$ is an orthogonal basis of $L^2(\{-1,1\}^p)$ }:

     $ \begin{aligned}[t]
      f&= \sum_{J\in [p]} \langle f, w_J \rangle w_J \\
      &= \sum_{J\in [p], |J| \leq d} \langle f, w_J \rangle w_J + \sum_{J\in [p], |J| > d} \langle f, w_J \rangle w_J\\
      &= f^{\leq d} + f^{>d}
      \end{aligned} $
    \item Marginal decomposition for multivariate densities on Boolean cube:
    \begin{align*}
        &\forall d\in \mathbb{N}^* ,\:\forall J \subset [p]\:; |J| = d,\: (\varepsilon_i)_{i\in J} \in \{-1,1\}^J,\\
        & \;\exists (\alpha_{J'})_{J'\subseteq J}\;;\\
        &\sum_{x \in\{-1,1\}^p}f(x)\boldsymbol{1}_{\{\forall i \in J, x(i) = \varepsilon_i\}} = \sum_{J' \subseteq J} \alpha_{J'}\langle f, w_{J'} \rangle(x)
    \end{align*}
\end{itemize}

These properties mean that each d-dimensional marginal of a random variable on the Boolean cube is entirely described by $f^{\leq d}$, the "low-frequency part" of $f$. We now introduce the algorithm \ref{alg:example} of private Sampling taken from \cite{PS}.

\begin{algorithm}[tb]
   \caption{Private sampling synthetic data algorithm}
   \label{alg:example}
\begin{algorithmic}
   \STATE {\bfseries Input:} a sequence $X$ of $n$ points in $\{-1, 1\}^p$ (true data); $m$ : cardinality of $S$; $d$ : the degree of the marginals to be matched; parameters $\delta$,$\Delta$ with $\Delta > \delta > 0$.
   \STATE 1. Draw a sequence $S = (\theta_1, . . . , \theta_m)$ of $m$ points in the cube independently and uniformly (reduced space).
   \STATE 2. Form the $m \times \binom{p}{ \leq d}$ matrix $M$ with entries $w_J$ $(\theta_i)$, i.e. the matrix whose rows are indexed by the points of the reduced space $S$ and whose columns are indexed by the Walsh functions of degree at most $d$. If the smallest singular value of $M$ is bounded below by $\sqrt{m}/2e^d$ call $S$ well conditioned and proceed. Otherwise, return “Failure” and stop.
   \STATE 3. Let $f_n$ be the uniform density on true data:
   
   $f_n = \frac{1}{n}\sum_{i=1}^{n}\boldsymbol{1}_{x_i}$.
   Consider the solution space : 
   $\begin{aligned}[t]
   H &= H(f_n) \\
   &= \left\{ h : \{-1, 1\}^p \to \mathbb{R} : supp(h) \subset S, h^{\leq d} = (f_n)^{\leq d}\right\}
    \end{aligned}$
   \STATE 4. Shrink $H$ toward the uniform density on S:
   \[u_m = \frac{1}{m}\sum_{i=1}^m \boldsymbol{1}_{s_i}\]  Let \[\tilde{H}=(1-\lambda)H+\lambda u_m\] where $\lambda \in [0, 1]$ is the minimal number such that \[\tilde{H} \cap [2\delta/m,(\Delta-\delta)/m]^S \neq \emptyset\]
   \STATE 5. Pick a proximal point :
  \[h^* = \argmin \left\{ \|\tilde{h}-u_m\|_{L^2(S)} : \tilde{h} \in \tilde{H} \cap[\delta/m,\Delta/m]^S\right\}\]
   \STATE {\bfseries Output:} a sequence $Y = (y_1, . . . , y_k)$ of $k$ independent points drawn from $S$ according to density $h^*$.
   
\end{algorithmic}
\end{algorithm}

\section{Optimization for private sampling}

In this section, we provide the two optimizations performed in order to compute $\lambda$ and $h^*$. These optimizations are used for the steps 4 and 5 of the algorithm \ref{alg:example}.
\newpage
\subsection{Fitting $\lambda$}
Step 4 of the algorithm \ref{alg:example} requires searching a minima. We take the notation of algorithm \ref{alg:example}. The problem is the following : 
 \[ \underset{\tilde{H} \cap [2\delta/m,(\Delta-\delta)/m]^S \neq \emptyset}{min} \lambda\]
Let $\tilde{h} \in \tilde{H} \cap [2\delta/m,(\Delta-\delta)/m]^S$. 
Then,
\begin{equation}
    \begin{cases}
      supp((\tilde{h}-\lambda u_m)/(1-\lambda))\subset S\\
      ((\tilde{h}-\lambda u_m)/(1-\lambda))^{\leq d} = f^{\leq d}\\
      \forall x \in \{-1,1\}^p, \tilde{h}(x) \in [2\delta/m,(\Delta-\delta)/m]\\
    \end{cases}\,
\end{equation}
Then we note $\hat{h} = (\lambda, h_1,...,h_m)^T$ the values of $\lambda$ and $\tilde{h}$ for each element of $S$, $W_S = (w_J(s))_{|J|\leq d, s \in S}$ and $W_X = (w_J(x))_{|J|\leq d, x \in X}$ the d-dimensional Walsh matrices on the spaces $S$ and $X$, $l = \binom{p}{ \leq d}$, $I_n = (1,...,1)^T \in \mathbb{R}^{1\times n}$,
\begin{align*}
    A_m, A_n &= (\boldsymbol{\delta}_{i,0})_{i,j} \in \mathbb{R}^{m \times m+1}, \mathbb{R}^{n \times m+1},\\
    B &= (\boldsymbol{\delta}_{i+1,j})_{i,j} \in \mathbb{R}^{m \times m+1}
\end{align*}
where $\boldsymbol{\delta}_{i,j}$ is the Kronecker delta symbol.\\\\
Then the system is :
\begin{align*}
    \lambda &=\underset{\hat{h}}{min}\;(1,0,...,0)^T \cdot \hat{h}\\
    &\begin{cases}
      \left(W_S  B  - \frac{1}{m}W_S A_m  + \frac{1}{n}W_x A_n\right) \cdot \hat{h} = \frac{1}{n}W_x I_n\\
      (0,2\delta/m,...,2\delta/m)^T \leq \hat{h}\\
      \hat{h} \leq (1,(\Delta-\delta)/m,...,(\Delta-\delta)/m)^T\\
    \end{cases}\,
\end{align*}

which is a linear program with $2|S|+l$ constraints.
\subsection{Fitting $\tilde{h}$}

Step 5 of algorithm \ref{alg:example} is another optimization. Taking the same notations as before, Step 5 can be reduced to a quadratic programming problem :
\begin{align*}
    h^* &=\underset{\tilde{h}}{\argmin} \; \tilde{h}^T \tilde{h} - 2 u_m^T \tilde{h}\\
    &\begin{cases}
      W_S \tilde{h} = \frac{1-\lambda}{n}W_x I_n + \frac{\lambda}{m}W_S I_m \\
      (\delta/m,...,\delta/m)^T \leq \tilde{h}\\
      \tilde{h} \leq (\Delta/m,...,\Delta/m)^T\\
    \end{cases}\,
\end{align*}

Which is a quadratic program with $2|S|+l$ constraints.

\section{Experimentations and discussion}
Private sampling introduces some constraints which must be respected in order to get differential privacy guarantees and control the accuracy \cite{PS}.
We also recall the two theorems providing bounds for differential privacy and accuracy \cite{PS}:
\begin{theorem}[Privacy]
If \[k \leq \frac{1}{4\sqrt{2}}\varepsilon(\delta/\Delta)^{3/2}e^{-d/2}\binom{p}{ \leq d}^{-1/4}\sqrt{n}/m^{3/4},\] then the algorithm \ref{alg:example} is $\varepsilon$-differentially private.
\end{theorem}
\begin{theorem}[Accuracy]
 Assume the true data $X = (x_1, . . . , x_n)$ is drawn independently from the cube according to some density f, which satisfies $\|f_n\|_\infty \leq \frac{\Delta}{2^p}$. Assume that 
 \begin{align*}
     n &\geq 16 \delta^{-2} \gamma^{-1}  e^{2d}\binom{p}{ \leq d}\\
     16 &\delta^{-2} \gamma^{-1} \Delta^2 e^{2d}\binom{p}{ \leq d} \leq m \leq 2^{p/4}\\
     k &\geq 4\delta^{-2}\left(log(2/\gamma)+log\binom{p}{ \leq d}\right)
 \end{align*}
 Then, with probability $1-4\gamma-\frac{1}{\sqrt{2^p}}$, the algorithm \ref{alg:example} succeeds, and all marginals of the synthetic data $Y$ up to dimension d are within $4\delta$ from the corresponding marginals of the true data.
\end{theorem}
Before getting to the results, we enumerate some of the assumptions made on $m$, $n$, $p$, $d$ and the density to get these two theorems :
\begin{equation}
    \begin{cases}
      \frac{16 \Delta^2 e^{2d}\binom{p}{ \leq d}}{\delta^{2} \gamma^{1}}   \leq m \leq 2^{p/4}\\
      k \leq \frac{\delta^{3/2}\sqrt{n}\varepsilon}{4\sqrt{2}\Delta^{3/2}e^{d/2}\binom{p}{ \leq d}^{1/4}m^{3/4}}
    \end{cases}\,
\end{equation}
 
These constraints seem too restrictive for a practical use. In order to show this, we take popular benchmark datasets \cite{UCI} for synthetic data generation, and we compute the required values for two-dimensional marginals matching ($d=2$ for all the discussion). These benchmark datasets are composed of categorial features. We one-hot encode each feature in order to get rows in the Boolean cube. Then, the assumptions on the data made for private sampling are verified. For the $Adult$ dataset, we take a smaller version of the dataset by removing all the numerical features. Here is the summary of the datasets :

\begin{table}[ht]
\caption{Summary of different standard benchmark datasets for private sampling}
\label{sample-table}
\vskip 0.15in
\begin{center}
\begin{small}
\begin{sc}
\begin{tabular}{lcccr}
\toprule
Dataset & $p$ & $n$ & $2^{p}$ & $\|f_n\|_\infty$\\
\midrule
Asia    & 8 & 20000 & 256 & $2.9 \cdot 10^{-1}$\\
Mushroom & 119 & 8124 & $6.7 \cdot 10^{35}$ & $1.2 \cdot 10^{-4}$\\
Car    & 25 & 1727 & $3.4 \cdot 10^{7}$ & $5.8 \cdot 10^{-4}$ \\
Adult   & 62 & 32561 &  $4.6 \cdot 10^{18}$ & $1.8 \cdot 10^{-2}$  \\
\bottomrule
\end{tabular}
\end{sc}
\end{small}
\end{center}
\label{tab:label}
\vskip -0.1in
\end{table}
The $Mushroom$ and $Car$ datasets correspond to an ideal case for private sampling. $\|f_n\|_\infty$ is low because each element of $\{-1,1\}^p$ in the dataset only appear once. Under the assumptions (2), theorem 4.2 ensures $4\delta$-accuracy for the computed density so $\delta \leq 1/4$. If we want to have at least $50\%$ of chances that the algorithm succeeds and have $\varepsilon$-DP, we get these relaxed constraints :
\begin{equation}
    \begin{cases}
      5.6 \cdot 10^{4} \cdot \|f_n\|_\infty^2 \cdot p(p+1) \cdot 2^{2p} \leq m \leq 2^{p/4}\\
      k \leq 9.7 \cdot 10^{-3} \cdot \frac{\sqrt{n}}{2^{3p/2}\|f_n\|_\infty ^{3/2}\cdot(p(p+1))^{1/4}m^{3/4}} \cdot  \varepsilon
    \end{cases}\,
\end{equation}
We compute these constraints for each dataset :
\begin{table}[ht]
\caption{Upper bounds for $k$ and $m$ and lower bounds for $m$ ,for $\varepsilon = 1$, $d=2$, $\delta \geq 1/4$ and $\gamma \leq 1/8$.}
\label{sample-table3}
\begin{center}
\begin{small}
\begin{sc}
\resizebox{\linewidth}{!}{
\begin{tabular}{lcccr}
\toprule
Dataset & $k$ : $upper$ $bound$ & $m$ : $lower$ $bound$ & $2^{p/4}$ \\
\midrule
Asia    & $7.3 \cdot 10^{-4} / m^{3/4}$ & $2.2 \cdot 10^{10}$ & 4 \\
Mushroom & $1.1 \cdot 10^{-49} / m^{3/4}$ & $5.3 \cdot 10^{72}$ & $9.0\cdot 10^{8}$\\
Car    & $2.9 \cdot 10^{-8}/ m^{3/4}$ & $1.4 \cdot 10^{16}$ & 77\\
Adult   & $9.5 \cdot 10^{-27}/ m^{3/4}$ & $1.5 \cdot 10^{42}$ &  46340\\
\bottomrule
\end{tabular}
}
\end{sc}
\end{small}
\end{center}
\label{tab:label2}
\vskip -0.1in
\end{table} 
 
 As Table \ref{tab:label2} shows, computing a density with private sampling may not be possible. In order to get differential privacy, the synthetic data must have a size which is close to zero. As $k$ is linear in $\varepsilon$, taking lower privacy constraints would not increase significantly $k$. Even if the size of the generated dataset was equal to 1 or a little bit greater, this algorithm does not enable to generate synthetic data for deep learning training from these datasets. This algorithm is closer to the query : "get one element of the dataset", which could be then differentially private.
 
 Moreover, the bounds on $m$ are inconsistent for these datasets and at the same time too high of a practical use of the algorithm. In fact, the lower bound on $m$ is often greater than $2^p$. In conclusion, for most standard datasets, the algorithm \ref{alg:example} is not applicable. Is there a situation where this algorithm could be used ?
 
 In order to answer this question, we place ourselves in a ideal case. We consider the dataset $X \supset \{-1,1\}^p$ such that $\|f_n\|_\infty = \frac{1}{2^p}$ is minimized. In this dataset, each element of $\{-1,1\}^p$ appear the exact same number of times. In this case, no shrinkage is performed as the density is regular. Then, the relaxed constraints become :
 \begin{equation}
    \begin{cases}
      5.6 \cdot 10^{4}  \cdot p(p+1) \leq m \leq 2^{p/4}\\
      k \leq 9.7 \cdot 10^{-3} \cdot \frac{\sqrt{n}}{(p(p+1))^{1/4}m^{3/4}} \cdot  \varepsilon
    \end{cases}\,
\end{equation}
We concentrate on the constraints on $k$. We have  $n \geq 2^p$ and, in the case $n = 2^p$:
\[k \leq 2.7 \cdot 10^{-6} \cdot \frac{2^{p/2}}{p(p+1)}\cdot \varepsilon\]
What would be the constraints on $k$ if we want to generate a dataset of size $1$, $100$, $1000$, $10000$ ? For multiple values of $\varepsilon$, we compute upper bounds for $k$ as a function of $p$ in order to get $\varepsilon$-differential privacy :
\begin{figure}[ht]
\vskip 0.2in
\begin{center}
\centerline{\includegraphics[width=\columnwidth]{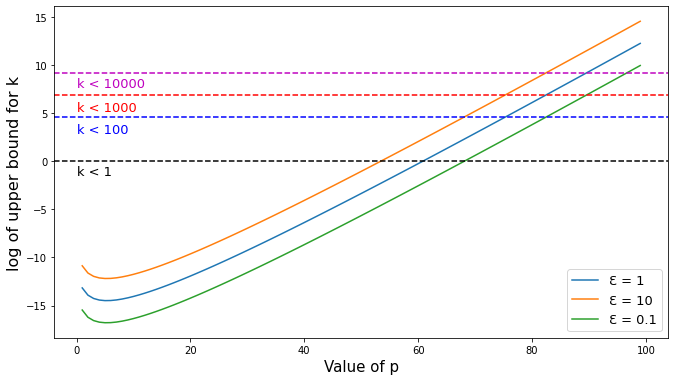}}
\caption{Log of the value of the upper bound for $k$ as a function of $p$ for $\varepsilon \in \{0.1,1,10\}$. Horizontal lines represent the thresholds for different $k \in \{1,100,1000,10000\}$. }
\label{icml-historical}
\end{center}
\vskip -0.2in
\end{figure}

Now that we have lower bounds on $p$ for $k$, we look at the lower bounds for $m$. The main objective is to assert if private sampling is applicable in this ideal case, i.e. if lower and higher bounds on $m$ defined by the theorem 4.1. can be respected and what would be the computational complexity of the algorithm \ref{alg:example}. In order to have the most relaxed constraints, we take $\varepsilon = 10$ for the numerical application :

\begin{table}[ht]
\caption{Lower bounds for $p$ and $m$ and higher bounds for $m$ for multiple sampling sizes and $\varepsilon = 10$.}
\label{sample-table2}
\vskip 0.15in
\begin{center}
\begin{small}
\begin{sc}
\begin{tabular}{lcccr}
\toprule
$k$  & $p$ : $lower$ $bound$ & $m$ : $lower$ $bound$ & $2^{p/4}$ \\
\midrule
1   & 46 & $1.2 \cdot 10^{8}$ & $7.0 \cdot 10^{13}$\\
100 & 68 & $2.7 \cdot 10^{8}$ & $3.0\cdot 10^{20}$\\
1000   & 75 & $3.2 \cdot 10^{8}$ & $3.8 \cdot 10^{22}$\\
10000 & 83 & $3.9 \cdot 10^{8}$ &  $9.7 \cdot 10^{24}$\\
\bottomrule
\end{tabular}
\end{sc}
\end{small}
\end{center}
\vskip -0.1in
\end{table}
With these values for $m$ and $p$, are the densities computationally tractable ? To answer this question, we compute a lower bound for computation time. A fast QP package, $quadprog$, has a complexity of at least $m^2$ for random dense problems. A random dense problem of size $1000$ is on average computed in 0.1s with $quadprog$ with an Intel i7-12700H. Then, a problem of size at least $1.2\cdot 10^{8}$ will be solved in at least $1.2\cdot10^9s$ which is 38 years. Then, even in the Boolean cube dataset case, with the most relaxed constraints and with a large $\varepsilon$, sampling of even one element with the algorithm \ref{alg:example} is not a computationally tractable program.

What if we took $n = l \cdot 2^p$ with $l > 1$ ? If we consider that private sampling is computationally tractable if the QP terminates in less than 1 month with an Intel i7-12700H, following the same reasoning and computing lower bounds on $m$ for $p$ from (4), we find $p \leq 7$. In this case, we have $5.6 \cdot 10^{4}  \cdot p(p+1) > 2^{p/4}$ which is not consistent and computing minimal values for $n$ from (4) with $k=1$, we find $n \in \Iintv{3 \cdot 10^{12},9 \cdot 10^{27}} $ depending on the value of p which is not a realistic size for a dataset.

\section{Conclusion}
In conclusion, private sampling is a method which comes with great premises. This algorithm generates a synthetic dataset in a polynomial time : the distribution is obtained after resolving a constrained quadratic program and a constrained linear programming problem. The synthetic data is also $\varepsilon$-differentially private, with $\varepsilon$ computable and accurate with an accuracy parameter  $\delta$ which can be tuned.

However, the constraints made on the distribution and the dataset make the algorithm impractical and ill-defined for real-world datasets. Even in ideal cases, the size of the quadratic program would be too big to for the algorithm to be computationally tractable.
\bibliographystyle{apacite}
\bibliography{main}

\begin{thebibliography}{}

\bibitem [\protect \citeauthoryear {%
Boedihardjo%
, Strohmer%
\BCBL {}\ \BBA {} Vershynin%
}{%
Boedihardjo%
\ \protect \BOthers {.}}{%
{\protect \APACyear {2021}}%
}]{%
PS}
\APACinsertmetastar {%
PS}%
\begin{APACrefauthors}%
Boedihardjo, M.%
, Strohmer, T.%
\BCBL {}\ \BBA {} Vershynin, R.%
\end{APACrefauthors}%
\unskip\
\newblock
\APACrefYearMonthDay{2021}{}{}.
\newblock
\APACrefbtitle {Private sampling: a noiseless approach for generating
  differentially private synthetic data.} {Private sampling: a noiseless
  approach for generating differentially private synthetic data.}
\PrintBackRefs{\CurrentBib}

\bibitem [\protect \citeauthoryear {%
Commission%
\ \BBA {} for Communication%
}{%
Commission%
\ \BBA {} for Communication%
}{%
{\protect \APACyear {2022}}%
}]{%
dataAct}
\APACinsertmetastar {%
dataAct}%
\begin{APACrefauthors}%
Commission, E.%
\BCBT {}\ \BBA {} for Communication, D\BHBI G.%
\end{APACrefauthors}%
\unskip\
\newblock
\APACrefYear{2022}.
\newblock
\APACrefbtitle {Data Act – The path to the digital decade} {Data act – the
  path to the digital decade}.
\newblock
\APACaddressPublisher{}{Publications Office of the European Union}.
\newblock
\begin{APACrefDOI} \doi{doi/10.2775/98413} \end{APACrefDOI}
\PrintBackRefs{\CurrentBib}

\bibitem [\protect \citeauthoryear {%
Dwork%
\ \BBA {} Roth%
}{%
Dwork%
\ \BBA {} Roth%
}{%
{\protect \APACyear {2014}}%
}]{%
DP}
\APACinsertmetastar {%
DP}%
\begin{APACrefauthors}%
Dwork, C.%
\BCBT {}\ \BBA {} Roth, A.%
\end{APACrefauthors}%
\unskip\
\newblock
\APACrefYearMonthDay{2014}{}{}.
\newblock
{\BBOQ}\APACrefatitle {The Algorithmic Foundations of Differential Privacy.}
  {The algorithmic foundations of differential privacy.}{\BBCQ}
\newblock
\APACjournalVolNumPages{Foundations and Trends in Theoretical Computer
  Science}{9}{3-4}{211-407}.
\newblock
\begin{APACrefURL}
  \url{http://dblp.uni-trier.de/db/journals/fttcs/fttcs9.html#DworkR14}
  \end{APACrefURL}
\PrintBackRefs{\CurrentBib}

\bibitem [\protect \citeauthoryear {%
Frank%
}{%
Frank%
}{%
{\protect \APACyear {2010}}%
}]{%
UCI}
\APACinsertmetastar {%
UCI}%
\begin{APACrefauthors}%
Frank, A.%
\end{APACrefauthors}%
\unskip\
\newblock
\APACrefYearMonthDay{2010}{}{}.
\newblock
{\BBOQ}\APACrefatitle {UCI machine learning repository} {Uci machine learning
  repository}.{\BBCQ}
\newblock
\APACjournalVolNumPages{http://archive. ics. uci. edu/ml}{}{}{}.
\PrintBackRefs{\CurrentBib}

\bibitem [\protect \citeauthoryear {%
Hayes%
, Melis%
, Danezis%
\BCBL {}\ \BBA {} Cristofaro%
}{%
Hayes%
\ \protect \BOthers {.}}{%
{\protect \APACyear {2018}}%
}]{%
GANINF}
\APACinsertmetastar {%
GANINF}%
\begin{APACrefauthors}%
Hayes, J.%
, Melis, L.%
, Danezis, G.%
\BCBL {}\ \BBA {} Cristofaro, E\BPBI D.%
\end{APACrefauthors}%
\unskip\
\newblock
\APACrefYearMonthDay{2018}{}{}.
\newblock
\APACrefbtitle {LOGAN: Membership Inference Attacks Against Generative Models.}
  {Logan: Membership inference attacks against generative models.}
\PrintBackRefs{\CurrentBib}

\bibitem [\protect \citeauthoryear {%
Lee%
, Kim%
, Jeong%
\BCBL {}\ \BBA {} Ro%
}{%
Lee%
\ \protect \BOthers {.}}{%
{\protect \APACyear {2022}}%
}]{%
NF}
\APACinsertmetastar {%
NF}%
\begin{APACrefauthors}%
Lee, J.%
, Kim, M.%
, Jeong, Y.%
\BCBL {}\ \BBA {} Ro, Y.%
\end{APACrefauthors}%
\unskip\
\newblock
\APACrefYearMonthDay{2022}{Jun.}{}.
\newblock
{\BBOQ}\APACrefatitle {Differentially Private Normalizing Flows for Synthetic
  Tabular Data Generation} {Differentially private normalizing flows for
  synthetic tabular data generation}.{\BBCQ}
\newblock
\APACjournalVolNumPages{Proceedings of the AAAI Conference on Artificial
  Intelligence}{36}{7}{7345-7353}.
\newblock
\begin{APACrefURL} \url{https://ojs.aaai.org/index.php/AAAI/article/view/20697}
  \end{APACrefURL}
\newblock
\begin{APACrefDOI} \doi{10.1609/aaai.v36i7.20697} \end{APACrefDOI}
\PrintBackRefs{\CurrentBib}

\bibitem [\protect \citeauthoryear {%
McKenna%
, Miklau%
\BCBL {}\ \BBA {} Sheldon%
}{%
McKenna%
\ \protect \BOthers {.}}{%
{\protect \APACyear {2021}}%
}]{%
NIST}
\APACinsertmetastar {%
NIST}%
\begin{APACrefauthors}%
McKenna, R.%
, Miklau, G.%
\BCBL {}\ \BBA {} Sheldon, D.%
\end{APACrefauthors}%
\unskip\
\newblock
\APACrefYearMonthDay{2021}{}{}.
\newblock
\APACrefbtitle {Winning the NIST Contest: A scalable and general approach to
  differentially private synthetic data.} {Winning the nist contest: A scalable
  and general approach to differentially private synthetic data.}
\PrintBackRefs{\CurrentBib}

\bibitem [\protect \citeauthoryear {%
Samarati%
\ \BBA {} Sweeney%
}{%
Samarati%
\ \BBA {} Sweeney%
}{%
{\protect \APACyear {1998}}%
}]{%
KANO}
\APACinsertmetastar {%
KANO}%
\begin{APACrefauthors}%
Samarati, P.%
\BCBT {}\ \BBA {} Sweeney, L.%
\end{APACrefauthors}%
\unskip\
\newblock
\APACrefYearMonthDay{1998}{}{}.
\newblock
{\BBOQ}\APACrefatitle {Protecting privacy when disclosing information:
  k-anonymity and its enforcement through generalization and suppression}
  {Protecting privacy when disclosing information: k-anonymity and its
  enforcement through generalization and suppression}.{\BBCQ}.
\newblock
\begin{APACrefURL} \url{https://api.semanticscholar.org/CorpusID:2181340}
  \end{APACrefURL}
\PrintBackRefs{\CurrentBib}

\end{thebibliography}

\end{document}